\documentclass[aps,notitlepage, longbibliography]{revtex4-1}

\usepackage{mathtools}
\usepackage{amsmath}
\usepackage{amssymb}
\usepackage[margin=1in]{geometry}
\usepackage{color}
\usepackage{graphicx}
\usepackage{multirow}
\usepackage{natbib}
\usepackage{mathrsfs}
\usepackage{mathtools}
\usepackage{lipsum}
\usepackage{lineno}
\usepackage[makeroom]{cancel}
\usepackage{comment}
\newtheorem{theorem}{Theorem}

\usepackage[normalem]{ulem}

\usepackage[dvipsnames]{xcolor}
\usepackage[colorlinks=true,
            linkcolor=blue,
            urlcolor=blue,
            citecolor=blue]{hyperref}

\usepackage{todonotes}

\begin{document}

\newcommand{\ovl}[1]{\overline{#1}}
\newcommand{\br}[1]{\left( #1 \right)}
\newcommand{\DR}{\Delta R_0}
\newcommand{\DP}{\Delta P_0}
\newcommand{\DA}{\Delta A}
\newcommand{\DB}{\Delta B}
\newcommand{\com}[1]{{\color{cyan}[#1]}}

\title{A {\it priori} screening of data-enabled turbulence models}

\author{Peng E S Chen$^{1}$, Yuanwei Bin$^{2,3,6}$, Xiang I A Yang$^2\footnote{Email: xzy48@psu.edu}$, Yipeng Shi$^{3,6}$, Mahdi Abkar$^4$, and George I. Park$^5$ \\
{\small $^1$Department of Mechanics and Aerospace Engineering, Southern University of Science and Technology, Shenzhen, China, 518055\\
$^2$Mechanical Engineering, Pennsylvania State University, PA, USA, 16802\\
$^3$State Key Laboratory of Turbulence and Complex Systems, Beijing, China, 100871\\
$^4$ Department of Mechanical and Production Engineering, Aarhus University,  Aarhus N, Denmark, 8200\\
$^5$Department of Mechanical Engineering and Applied Mechanics, University of Pennsylvania, Philadelphia, PA, USA, 19104\\
$^6$College of Engineering, Peking University, China, 100871\\
}
}

\begin{abstract}
Assessing the compliance of a white-box turbulence model with known turbulent knowledge is straightforward.
It enables users to screen conventional turbulence models and identify apparent inadequacies, thereby allowing for a more focused and fruitful validation and verification.
However, comparing a black-box machine-learning model to known empirical scalings is not straightforward.
Unless one implements and tests the model, it would not be clear if a machine-learning model, trained at finite Reynolds numbers preserves the {\it known} high Reynolds number limit.
This is inconvenient, particularly because model implementation involves retraining and re-interfacing. 
This work attempts to address this issue, allowing fast {\it a priori} screening of machine-learning models that are based on feed-forward neural networks (FNN).
The method leverages the mathematical theorems we present in the paper.
These theorems offer estimates of a network's limits even when the exact weights and biases are unknown.
For demonstration purposes, we screen existing machine-learning wall models and RANS models for their compliance with the log layer physics and the viscous layer physics in an {\it a priori} manner.
In addition, the theorems serve as essential guidelines for future machine-learning models.
\end{abstract}

\maketitle

\section{Introduction}
\label{sec:intro}

The range of scales in high Reynolds number turbulent flows spans multiple orders of magnitude. 
Conducting a direct numerical simulation (DNS) that resolves all these scales is prohibitively expensive at high Reynolds numbers \cite{li2022grid, yang2021grid, choi2012grid}, leading to the need for turbulence modeling. 
Examples of turbulence modeling include sub-grid scale modeling and wall modeling in large-eddy simulation (LES) \cite{meneveau2000scale, piomelli2002wall, bose2018wall}, as well as Reynolds stress modeling in Reynolds-averaged Navier-Stokes (RANS) \cite{durbin2018some}.
Conventional turbulence models (in particular, RANS models) rely heavily on empirical scalings, in addition to knowledge derived from first principles such as Galilean invariance and realizability. 
These empirical scalings include power-law decay of unforced isotropic turbulence \cite{thormann2014decay}, the law of the wall (LoW) \cite{marusic2013logarithmic}, Kolmogorov's hypotheses of small-scale turbulence \cite{kolmogorov1941dissipation}, as well as Townsend's attached eddy hypothesis \cite{marusic2019attached,yang2019hierarchical,yang2018hierarchical, duraisamy2023uncovering}, among many others \cite{pope2000turbulent}. 
While these empirical scalings are not derived from first principles, they have been extensively validated and are expected to hold even under unseen conditions.
The most well-known example is probably the LoW.
Although it cannot be derived directly from the Navier-Stokes equations, numerous studies have demonstrated its validity in canonical wall flows (pipe, channel, flat plate) with no bulk acceleration 
\cite{hutchins2009hot, hultmark2012turbulent, rosenberg2013turbulence,lee2015direct, hoyas2022wall, xu2021flow}.
It is commonly believed that the log law remains valid at larger Reynolds numbers, potentially extending to infinity.
Compliance with empiricisms like the log law serves as a straightforward criterion for assessing turbulence models. 
For instance, a model intended for wall-bounded flows but failing to preserve the law of the wall is deficient and should be discarded \cite{spalart2015philosophies}. 
Being able to confidently screen out a model with critical flaws without having to implement/verify/validate it is not only prudent and robust from the physics point of view, but also efficient from a user's perspective \cite{duraisamy2019turbulence, pradhan2023unified}: V\&V campaigns for RANS models require considerable time and collective efforts \cite{rumsey2011summary, rumsey2015overview, rumsey2019overview, tinoco2018summary}.

A need for such simple criteria exists for data-based machine-learning models as well.
In fact, with the increasing number of machine-learning models available \cite{singh2017machine,tenney2020application,yin2022iterative,fang2023toward,pan2018data,wang2017physics,zhao2020rans,xie2021artificial,xie2020modeling,huang2019wall,Vadrot2023} and the challenges associated with their implementations \cite{rumsey2022search,vadrot2022survey}, the ability to identify models that are lacking extrapolation of key flow physics 
\color{black}
is more valuable for machine-learning models than for conventional empirical models.
However, assessing whether a machine-learning model adheres to known knowledge is not a straightforward task. 
Unlike conventional turbulence models that are white-box models, machine-learning models,
except for a few exceptions \cite{weatheritt2017development, zhao2020rans, brunton2016discovering,hansen2023data}, are black-box ones.
How to evaluate the asymptotic behavior of a black box in an {\it a priori} manner is not clear. 
Consider, for example, the machine-learning model in Ref. \cite{yang2019predictive}.
The model is a feedforward neural network. 
The network is trained against the $\rm Re_\tau=1000$ channel flow DNS in Ref. \cite{graham2016web}.
It takes the instantaneous $U^+/\ln (y^+)$ and $y^+/U^+$ as its inputs and computes the instantaneous wall-shear stress $\tau^+_w$ as its output.
It is not evident in an {\it a priori} sense whether this trained network would preserve \color{black} 
the law of the wall.
As a result, it is difficult to dismiss a machine-learning model without implementing, validating, and verifying it. 
This is undesirable, particularly considering that implementing a machine-learning model involves re-training and adapting it to different code environments.
Furthermore, due to the lack of satisfactory results from machine-learning models in the field of computational fluid dynamics (CFD) \cite{rumsey2022search}, the high labor cost associated with assessing these models hinders their widespread adoption in engineering practice.

The present work \color{black} aims to tackle the aforementioned challenge. 
We focus on a priori examination of whether feedforward-neural-network-based (FNN-based) machine-learning models respect the LoW. 
While it is acknowledged that not all machine-learning models are based on FNNs \cite{bhatnagar2019prediction, xu2020multi, duvall2021discretization, bakarji2022dimensionally, liu2022new, huang2023distilling,xiang2021neuroevolution, li2023long}, a substantial number of them are. 
Examples include tensor-based neural networks \cite{ling2016reynolds}, physics-informed machine learning \cite{xiao2020flows, tao2020physics}, field inversion and machine learning \cite{singh2017machine}, progressive machine learning \cite{bin2022progressive,bin2023data}, among others. 
Regarding the LoW, although there is other known knowledge that is fundamental to turbulence models, the LoW is regarded by many to be the most important one \cite{menter2019best, spalart2015philosophies}. 
The remainder of the paper is organized as follows.
We present the technical background and the mathematical theorems in section \ref{sec:theory}.
In section \ref{sec:result}, we apply the theorems to assess the existing machine-learning models in the literature.
Finally, we conclude in section \ref{sec:conclusions}.

\section{Background and theorems}
\label{sec:theory}

This section is organized as follows.
An overview of FNN is given in section \ref{sub:FNN}.
In section \ref{sub:theorem}, we present four mathematical theorems which form the foundation of the present work. 
Their proofs are presented in the Appendix.
Lastly, we summarize the physical scalings in section \ref{sub:empi}.

\subsection{Feedforward neural networks}
\label{sub:FNN}

\begin{figure}
    \centering
    \includegraphics[width=0.4\textwidth]{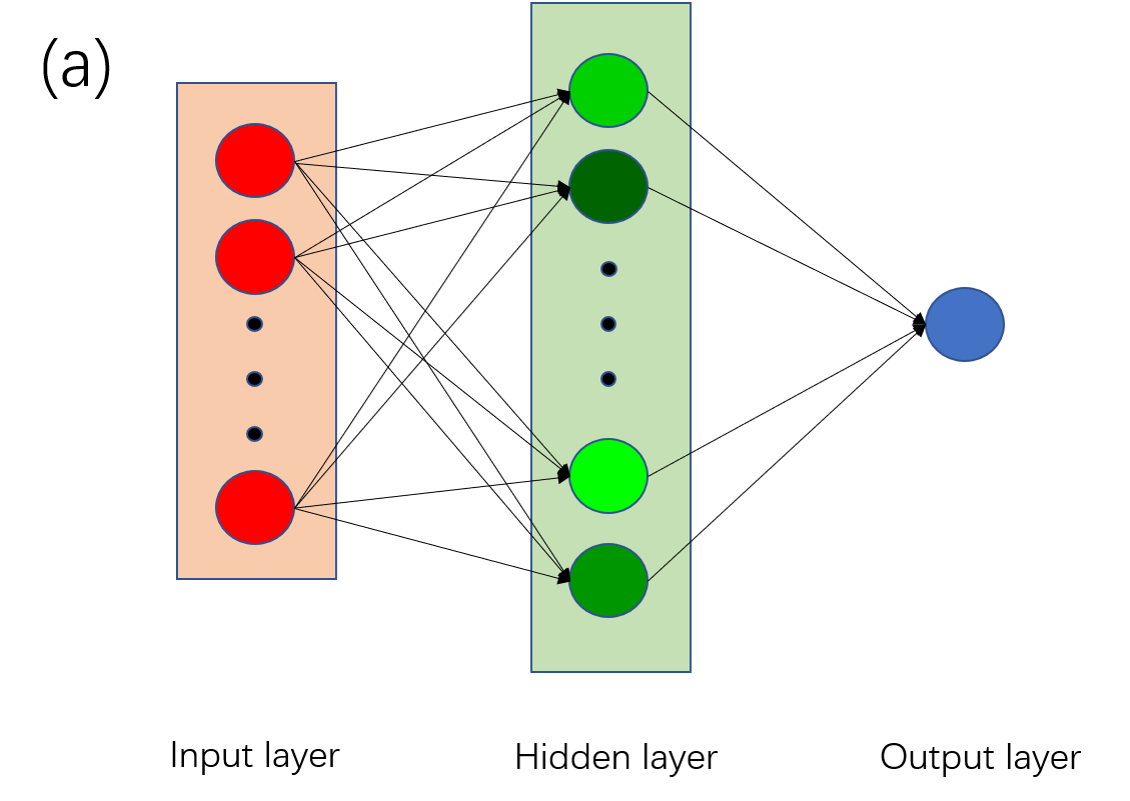}
    \includegraphics[width=0.36\textwidth]{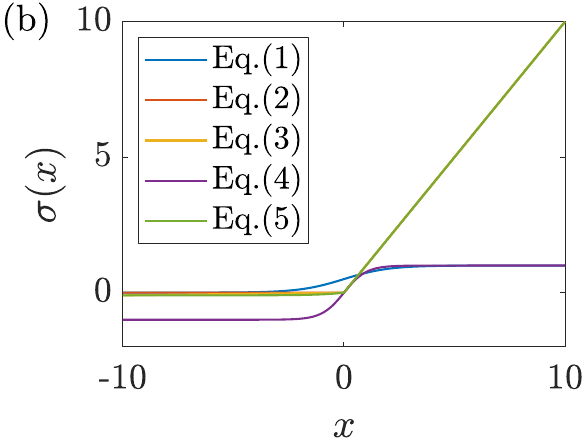}
    \caption{(a) A schematic of an FNN. The shade of the color represents the value of the activation function. (b) Some activation functions.}
    \label{fig:sketch}
\end{figure}

Figure \ref{fig:sketch}(a) shows a schematic of a feedforward neural network (FNN).
The FNN contains an input layer, a hidden layer, and an output layer.
A general FNN maps $\mathbb{R}^N$ to $\mathbb{R}^M$, where $N$ and $M$ are the dimensions of the input and the output space. 
Such an FNN can, however, be split into $M$ FNNs whose output is strictly one-dimensional.
Without loss of generality, we focus on the case where $M=1$.

The neurons in the hidden layer receive input signals from the previous layer, compute outputs, and send the output signals to the next layer. 
The commonly employed activation functions include the sigmoid function, the ReLU function, the leaky-ReLU function, the tanh-sigmoid function, and the exponential linear unit function, which are ploted in figure \ref{fig:sketch}(b) and are shown below.
The sigmoid function reads
\begin{equation}
    \sigma(x)=\frac{1}{1+e^{-x}}.
    \label{eq:sigmoid}
\end{equation}
The ReLU function reads
\begin{equation}
    \sigma(x)=\begin{cases}
        x & x\ge 0\\
        0 & x<0
    \end{cases}.
    \label{eq:relu}
\end{equation}
The leaky-ReLU function reads
\begin{equation}
    \sigma(x)=
    \begin{cases}
        x & x\ge 0\\
        \gamma x & x<0
    \end{cases},
    \label{eq:relu_leak}
\end{equation}
where $\gamma$ is a parameter.
The tanh-sigmoid function reads
\begin{equation}
\sigma(x)=\frac{1-e^{-2x}}{1+e^{-2x}}.
\label{eq:tan_sigmoid}
\end{equation}
The exponential linear unit reads
\begin{equation}
    \sigma(x)=
    \begin{cases}
    x & \text{, if } x\ge 0 \\
    a(e^x-1) & \text{, otherwise}.
    \end{cases}
    \label{eq:exp_linear}
\end{equation}
Table \ref{tab:asy_act} tabulates the properties of these activation functions.
An activation function, which is a map from $\mathbb{R}$ to $\mathbb{R}$ (i.e., $\sigma: \mathbb{R}\to\mathbb{R}$), is bounded if there is a real $c$ such that for any $x\in  \mathbb{R}$, $|\sigma(x)|<c$. Otherwise, the activation function is called unbounded.

\begin{table}
    \centering
    \begin{tabular}{lccc}\hline
         Name &  $x\to +\infty$ & $x\to -\infty$  \\\hline
         sigmoid &  1 & 0\\
         ReLU &  $O(x)$  & 0 \\
         leaky-ReLU  & $O(x)$  & $O(x)$ \\
         tanh-sigmoid & 1 & -1 \\
         Exponential linear unit  & $O(x)$ & 0\\
         \hline
    \end{tabular}
    \caption{Properties of the activation functions.}
    \label{tab:asy_act}
\end{table}

\subsection{Theorems and their implications}
\label{sub:theorem}

The discussion here pertains only to the 
commonly used the activation functions in Table \ref{tab:asy_act}, 
as well as the FNNs whose output space is one-dimensional as discussed in the previous subsection. 
\color{black}
We also assume that the activation functions of the neurons within a hidden layer are identical. 
This is not restrictive, since most FNNs utilize the same activation function for all neurons.
The following theorems are reminiscent of the extrapolation theorem in Ref. \cite{bin2022progressive},  but they are stronger (in terms of their mathematical properties). 

\begin{theorem}
Consider an FNN with one hidden layer:
if its activation function is bounded, its output is also bounded.  
\label{th:singlebound}
\end{theorem}
Hereinafter, $f$ denotes an FNN, and $\sigma$ denotes the activation function.

\begin{theorem}
Consider a multilayer FNN that contains $r$ hidden layers with $\sigma_i(x)$, $i=1,2,...,r$ being the activation functions of the neurons in the $r$th hidden layers: if one of the $r$ activation functions is bounded, the multilayer FNN is bounded.
\label{th:multibound}
\end{theorem}

\begin{theorem}
Consider an FNN with one hidden layer:
if  ~$\sigma(x)=O(x^\alpha)$, as $x\to \pm\infty$, $\alpha>0$, then $f(\mathbf{x})= O(\max\limits_{1\le i \le N} |x_i|^\alpha)$ as $\max\limits_{1\le i \le N} |x_i|\to \infty$.
\label{th:singleunbound}
\end{theorem}
Here, $O(\cdot)$ is ``on the order of''.

\begin{theorem}
Consider a multilayer FNN that contains $r$ hidden layers with $\sigma_i(x)$, $i=1,2,...,r$ being the activation functions of the neurons in the $r^{th}$ hidden layers:
if ~$\sigma_i(x)=O(x^{\alpha_i})$ as $x\to \infty$, $\alpha_i>0$, $i=1,2,...,r$, then $f(\mathbf{x})=O((\max\limits_{1\le i \le N} |x_i|)^\alpha)$  as $\max\limits_{1\le i \le N} |x_i|\to \infty$, with $\alpha=\alpha_1\alpha_2\cdots\alpha_r$.
\label{th:multiunbound}
\end{theorem}
For the activation functions in Table \ref{tab:asy_act}, $\alpha$ in {Theorem 3} and $\alpha_i$ in {Theorem 4}  may only take values of 1 (ReLu when $x\to +\infty$, leaky-ReLu, Exponential linear unit when $x\to +\infty$) or 0 (sigmoid, tanh-sigmoid, ReLu when $x\to -\infty$, exponential linear unit $x\to -\infty$). 
Theorems \ref{th:singlebound} and \ref{th:multibound} establish that the output of a neural network employing either the sigmoid or tanh-sigmoid function is always bounded, regardless of whether the input is bounded or not. 
Theorems \ref{th:singleunbound} and \ref{th:multiunbound} present slightly stronger conclusions than their counterparts.
According to Theorems \ref{th:singleunbound} and \ref{th:multiunbound}, the growth rate of the activation function directly influences the rate at which a network's output increases at the limit of infinite.

These theorems characterize FNN's limiting behavior with unbounded inputs. In the present study, this is useful in understanding the FNN-based models' response at the limit of high Reynolds numbers (Reynolds number scaling). 

To briefly illustrate the usefulness of the Theorems, we take the Reynolds number scaling of the centerline velocity in a channel as an example.
Denoted the centerline velocity as $U^+_c$. It scales as $U^+_c=\ln(\rm Re_\tau)/\kappa$, when $\rm Re_\tau$ is large.
Here, $\rm Re_\tau$ represents the friction Reynolds number, and $\kappa$ corresponds to the von K{\'a}rm{\'a}n constant.
In light of Theorems \ref{th:singlebound} and \ref{th:multibound}, training an FNN with the sigmoid or tanh-sigmoid activation function to predict $U^+_c$ as a function of $\rm Re_\tau$ would not give a generalizable model due to the unbounded nature of the output.
Theorems \ref{th:singleunbound} and \ref{th:multiunbound} indicate that utilizing the remaining activation functions listed in Table \ref{tab:asy_act} would also be insufficient to predict $U^+$ as a function of $\rm Re_\tau$  because the activation functions mentioned in Table \ref{tab:asy_act} yield power-law scalings rather than logarithmic scalings at infinity.

\subsection{Physical knowledge}
\label{sub:empi}

In this subsection, we list the physics that a turbulence model should preserve.
First, the law of the wall dictates that the mean flow in the logarithmic layer scales as
\begin{equation}
U^+=\frac{1}{\kappa}\ln(y^+)+B, 
\end{equation}
where $U$ is the streamwise mean velocity, $y$ is the wall-normal coordinate, and  $B\approx 5.0$ is the log law intercept. The superscript $+$ denotes normalization by the inner units 
(kinematic viscosity $\nu$ and friction velocity $u_\tau$). 
At a given distance $h_{wm}/\delta$ from the wall, we have the following:
\begin{equation}
 h_{wm}^+= O({\rm Re}_\tau), ~~\ U^+_h= O(\ln({\rm Re}_\tau)),
 \end{equation}
as $\rm Re_\tau \to +\infty$, and the following:
 \begin{equation}
     ~~\lim_{{\rm Re}_\tau\to +\infty} \tau^+=1, ~~\lim_{{\rm Re}_\tau\to +\infty} \tau/(\rho_b U_b^2)=0, ~\text{and}~~ \lim_{\rm Re_\tau\to +\infty}\tau/(\rho_b U_h^2)=0.
\end{equation}
Here, $\delta$ is an outer length scale, $U_h$ is the fluid velocity at $h_{wm}$, $\tau$ is the wall-shear stress, $\rho_b$ is the bulk density, and $U_b$ is the bulk velocity (or some outer velocity scale).
Furthermore, the production approximately balances dissipation in the logarithmic layer. 

When the Reynolds number approaches 0, the flow approaches the laminar limit and we have 
\begin{equation}
U^+=y^+.
\end{equation}
The viscous length scale $\nu/u_\tau$ approaches infinity.
At a given distance $h_{wm}/\delta$ from the wall, we have
\begin{equation}
\lim_{{\rm Re}_\tau\to 0} h_{wm}^+=0,~~\lim_{{\rm Re}_\tau\to 0} U_{h}^+=0,~~\text{and}~ \lim_{{\rm Re}_\tau\to 0} U_h^+/h_{wm}^+=1.
\end{equation}

\section{Applications}
\label{sec:result}

We invoke the theorems outlined in \ref{sub:theorem} and screen the existing machine-learning models. 
We note that no criticism is implied when a model is found not to preserve the law of the wall.
Also, this section is not meant to be a comprehensive survey of all existing machine-learning models.
Section \ref{sub:wmles} focuses on LES wall models, and section \ref{sub:sgs}  focuses on RANS models.

\subsection{Wall-modeled LES}
\label{sub:wmles}

We study the data-based wall models in Refs. \cite{yang2019predictive, huang2019wall, dupuy2023data, zhou2021wall,zhou2023wall} and whether they preserve the log law at the high and low Reynolds number limits.
The details of these wall models are summarized in Tables \ref{tab:WMLES} and \ref{tab:WMLES-laminar} regarding their behaviors at the high and low Reynolds number limits, including their inputs, outputs, and the behaviors of these inputs and outputs as the Reynolds number approaches infinity and 0.
A distinction is made between the expected and the actual behaviors of the output: the former is due to the scalings in Section \ref{sub:empi} and is the desired behavior, whereas the latter is due to the FNN setup and is the actual behavior of the model.
We will use inline equations at most places.
Standalone equations are used only if the expressions are too long.

In Yang et al.\cite{yang2019predictive}, an FNN wall model utilizing the sigmoid activation function was trained. 
The model takes the following inputs: ${|u_\parallel^+|}/{h_{\rm \rm wm}^+}$ and ${\ln (h_{\rm \rm wm}^+/y_0^+)}/{|u_\parallel^+|}$, where $|u_\parallel^+|$ represents the instantaneous wall-parallel velocity at a distance $h_{\rm \rm wm}$ from the wall, and $y_0^+$ is a roughness/viscous scale.
For smooth walls, $y_0=\nu/u_\tau\exp(-\kappa B)$, and $y_0^+= O(1)$, as ${\rm Re}_\tau \to +\infty$.
{As ${\rm Re}_\tau$ approaches infinity, the inputs behave as follows
\begin{equation}
\begin{split}
    &\lim_{\rm Re_\tau\to +\infty}\frac{u^+_\parallel}{h_{\rm wm}^+}=\lim_{\rm Re_\tau\to +\infty} \frac{\ln( h_{\rm wm }^+)/\kappa+B}{h_{\rm wm}^+}=\lim_{\rm Re_\tau\to +\infty}\frac{\ln (\rm{Re}_\tau)/\kappa}{\rm{Re}_\tau}=0,\\
    &\lim_{\rm Re_\tau\to +\infty}\frac{\ln (h_{\rm \rm wm}^+/y_0^+)}{|u_\parallel^+|}=\lim_{\rm Re_\tau\to +\infty}\frac{\ln (\rm Re_\tau)}{\ln (\rm{Re}_\tau)/\kappa}=\kappa.
\end{split}
\end{equation}
The expected output $|\tau_{w,\parallel}^+|$ is finite. }
According to Theorem \ref{th:singlebound}, the actual output is finite.
Hence, this FNN, if well-trained, preserves the LoW at high Reynolds numbers. 
This aligns with the results reported in \cite{yang2019predictive}.
On the other hand, as $\rm Re_\tau$ approaches 0, the two inputs behave as follows
\begin{equation}
\begin{split}
&\lim_{\rm Re_\tau\to 0} \frac{u_\parallel^+}{h_{\rm wm}^+}=1, \\
&\lim_{\rm Re_\tau\to 0}\frac{\ln (h_{\rm \rm wm}^+/y_0^+)}{|u_\parallel^+|}=\lim_{h_{\rm wm}^+\to 0}\frac{\ln (h_{\rm wm}^+)}{h_{\rm wm}^+}\to -\infty.
\end{split}
\end{equation}
The expected and the actual outputs are, again, finite---thanks to the use of the bounded sigmoid activation function.
According to Theorem \ref{th:singlebound}, this FNN will also preserve the laminar limit.

In Huang et al.\cite{huang2019wall}, two FNNs were trained to predict the mean flow $U^+$ as a function of $y^+$ and a length scale related to system rotation, $l_\Omega^+$. 
In the absence of system rotation, $l_\Omega^+=\infty$ and $1/l_\Omega^+=0$ and only $y^+$ varies. 
The sigmoid activation function was employed for both FNNs. 
The first FNN takes $y^+$ and $1/l^+_\Omega (=0)$ as its inputs. 
As the Reynolds number approaches infinity, the non-zero input behaves as $y^+= O(\rm Re_\tau)$, and the expected output $U^+=O(\ln(\rm Re_\tau))$.
Per Theorem \ref{th:singlebound}, the actual output of the first FNN behaves as $U^+= O(1)$ at the infinite Reynolds number limit.
There is a misalignment between the expected and the actual outputs.
Consequently, the first FNN does not preserve the LoW at high Reynolds numbers, regardless of its training. 
The inputs of the second FNN are $y^+$ and $y^+/l^+_\Omega$, and the output is $U^+-g(y^+,l^+_\Omega)$, where 
\begin{equation}
    g(y^+,l_\Omega^+)=\frac{1}{\kappa}\ln(y^+)H\left(-y^++\frac{l_\Omega^+}{\kappa}\right)+\left[ \frac{y^+}{l_\Omega^+} + \frac{1}{\kappa}\ln (l_\Omega^+) -\frac{1}{\kappa}\ln (\kappa e) \right]H\left(y^+-\frac{l_\Omega^+}{\kappa}\right),
\end{equation}
$H$ is the Heaviside function and $e$ is the base of the natural logarithm. 
In the absence of system rotation, $g$ reduces to $g=\ln(y^+)/\kappa$.
When the Reynolds number approaches infinity, the non-zero input behaves as $y^+= O(\rm Re_\tau)$ and the expected output is 
$\lim_{\rm Re_\tau\to +\infty} (U^+-g(y^+,l_\Omega^+))=B$ and is finite.
The actual output according to Theorem \ref{th:singlebound} is also finite.
Hence, the second FNN preserves the LoW at high Reynolds numbers as the behavior of the actual output aligns with the behavior of the expected output.
{These findings are consistent with the results presented in Ref. \cite{yang2019predictive}.
There, it was observed that only the second network has the capability to extrapolate beyond the training data. 
We next consider the laminar limit.
When the Reynolds number approaches 0, both inputs of both FNNs tend to $0$.
The actual and expected outputs of the first FNN tends are finite, and therefore the first FNN preserves the laminar limit.
The expected output of the second FNN behaves
$\lim_{\rm Re_\tau\to 0} (U^+-g)=\lim_{y^+\to 0}(y^+-\ln(y^+)/\kappa)=+\infty$,
but the actual output is finite due to Theorem \ref{th:singlebound}.
Hence, the second FNN does not preserve the laminar limit.
}

In Dupuy et al.\cite{dupuy2023data}, an FNN wall model was trained using the exponential linear unit as the activation function. 
The inputs of the FNN are $h_{\rm wm} u_i/\nu=h_{\rm wm}^+ u_i^+$, where $i=1$, 2, and 3. The outputs of the FNN are $\rho(h_{\rm wm}/\mu)^2\tau_i=h_{\rm wm}^{+2}\tau^+$, where $\tau_i$ represents the instantaneous wall-shear stress in the $i$th direction. 
As $\rm Re_\tau $ approaches infinity, the inputs behave as ${h_{\rm wm}^+ u^+}=O(\rm{Re}_\tau\ln (\rm{Re}_\tau))$, and the expected output behaves as $\rho \left({h_{\rm wm}}/{\mu}\right)\tau=O( \rm Re_\tau^2)$.
According to Theorem \ref{th:multiunbound}, the actual output behaves as $O(Re_\tau\ln(Re_\tau))$.
We see a misalignment between the actual and the expected outputs.
Hence, the model does not preserve the law of the wall at high Reynolds numbers---regardless of its training.
In the limited of $\rm Re_\tau\to 0$,
the inputs behave as $\lim_{\rm Re_\tau\to 0}{h_{\rm wm} u}/{\nu}=\lim_{h_{wm}^+\to 0} h_{wm}^{+2}=0$, the expected and actual outputs are both finite.
Hence, the FNN, if properly trained, preserves the laminar limit.

Moving on to Zhou et al.\cite{zhou2021wall}, an FNN wall model was trained using the ReLU activation function. 
The inputs include $\ln(h_{\rm wm}/y^*)$, $u_i\delta/(U_bh_{\rm wm})$, and $(h_{\rm wm}/(\rho U_b^2))\partial_i p$. 
Here, $y^*=\nu/\sqrt{u_\tau^2+u_p^2}$, $u_\tau=\sqrt{\nu u_{||}/h_{\rm wm}}$, $u_p=(\nu/\rho \partial_x p)^{1/3}$, $u_i$ represents the velocity in the $i$th direction, $u_{||}$ represents the velocity in the wall-parallel direction, $\delta$ represents an outer length scale, and $\partial_i p$ represents the pressure gradient in the $i$th direction. 
{The discussion here is limited to the log layer, so that the pressure gradient effect can be ignored, and therefore we have $\partial_i p\approx 0$, $u_p\approx 0$, $y^*\approx \nu/u_\tau$.
When $\rm Re_\tau \to +\infty$,  the inputs behave as follows:
$\ln \left({h_{\rm wm}}/{y^*}\right)=O( \ln \rm Re_\tau)$,
${u \delta}/{U_b h_{\rm wm}}= O(1)$, and
${h_{\rm wm}}/{\rho U_b^2}\partial_i p\to 0$.
The expected output behaves as ${\tau_w}/{\rho U_b^2}= O(1)$.
According to Theorem \ref{th:multiunbound}, the actual output behaves as $O(\ln(Re_\tau))$.
There is a misalignment between the actual and the expected output, and therefore this FNN does not preserve the LoW at high Reynolds numbers.
Finally, in Zhou et al.\cite{zhou2023wall}, the FNN model shares the same input and output features as Zhou et al.\cite{zhou2021wall}, but it utilizes the tanh-sigmoid function as the activation function, which is bounded. 
As per Theorem \ref{th:singlebound}, the FNN preserves the LoW.
For the laminar limit, the behaviors are similar. 
According to Theorem \ref{th:multiunbound}, the FNN model presented in Ref. \cite{zhou2021wall} does not preserve the laminar flow limit, whereas the FNN model in Ref. \cite{zhou2023wall} does.

\begin{table}
\small
    \centering
    \scalebox{1}{\begin{tabular}{rccccccc}\hline
         Ref. & Inputs & Outputs & Activation & Input (Ex) & Output (Ex) & Output (Ac) & Test \\
         \hline
         \cite{yang2019predictive}& $\dfrac{|u_\parallel^+|}{h_{\rm \rm wm}^+}, \dfrac{\ln (h_{\rm \rm wm}^+/y_0^+)}{|u_\parallel^+|}$ & $|\tau^+_{w, \parallel}|$ & Eq. \eqref{eq:sigmoid} & F, F & F & F & Y\\
        \cite{huang2019wall} & $y^+$, $1/{l_\Omega^+}$ & $U^+$ & Eq. \eqref{eq:sigmoid} & $\rm Re_\tau$, F & $\ln \rm Re_\tau$ & F & N \\
        \cite{huang2019wall} & $y^+$, $y^+/l_\Omega^+$ & $U^+-g$ & Eq. \eqref{eq:sigmoid} & $\rm Re_\tau$, F & F & F & Y\\
        \cite{dupuy2023data} & $h_{\rm wm}u_i/\nu$ & $\rho(h_{\rm wm}/\mu)^2\tau$ & Eq. \eqref{eq:exp_linear} & $\rm Re_\tau\ln \rm Re_\tau $, F, F & $\rm Re_\tau^2$ & $\rm Re_\tau\ln \rm Re_\tau $& N\\
        \cite{zhou2021wall} & $\ln\left(\dfrac{h_{\rm wm}}{y^*}\right)$, $\dfrac{u_i}{U_b}\dfrac{\delta}{h_{\rm wm}}$, $\dfrac{h_{\rm wm}}{\rho U_b^2}\dfrac{\partial p}{\partial x_i}$ &$\dfrac{\tau_w}{\rho U_b^2}$ & Eq. \eqref{eq:relu} & $\ln \rm Re_\tau$, F, F & F & $\ln \rm Re_\tau$ & N\\
        \cite{zhou2023wall} & $\ln\left(\dfrac{h_{\rm wm}}{y^*}\right)$, $\dfrac{u_i}{U_b}\dfrac{\delta}{h_{\rm wm}}$, $\dfrac{h_{\rm wm}}{\rho U_b^2}\dfrac{\partial p}{\partial x_i}$ &$\dfrac{\tau_w}{\rho U_b^2}$ & Eq. \eqref{eq:tan_sigmoid} & $\ln \rm Re_\tau$, F, F & F & F & Y\\
    \hline
    \end{tabular}}
    \caption{Details of the FNN-based machine-learning wall models. The table lists the references, the inputs of the FNNs, the activation functions, the expected  (as indicated by ``Ex'') and the actual (indicated by ``Ac'') behaviors of the output as the Reynolds number approaches infinity. ``F'' indicates that the variable in question is finite. The last column indicates if the FNN has the potential to preserve the law of the wall at high Reynolds numbers with ``Y'' and ``N'' stand for ``yes'' and ``no''.}
    \label{tab:WMLES}
\end{table}

\color{blue}
\begin{table}
\small
    \centering
    \scalebox{1}{\begin{tabular}{rccccccc}\hline
         Ref. & Inputs & Outputs & Activation & Input (Ex) & Output (Ex) & Output (Ac) & Ability \\
         \hline
         \cite{yang2019predictive}& $\dfrac{|u_\parallel^+|}{h_{\rm \rm wm}^+}, \dfrac{\ln (h_{\rm \rm wm}^+/y_0^+)}{|u_\parallel^+|}$ & $|\tau^+_{w, \parallel}|$ & Eq. \eqref{eq:sigmoid} & F, $-\infty$ & F & F & Y\\
        \cite{huang2019wall} & $y^+$, $1/{l_\Omega^+}$ & $U^+$ & Eq. \eqref{eq:sigmoid} & F, F & F & F & Y \\
        \cite{huang2019wall} & $y^+$, $y^+/l_\Omega^+$ & $U^+-g$ & Eq. \eqref{eq:sigmoid} & F, F & $-\infty$ & F & N\\
        \cite{dupuy2023data} & $h_{\rm wm}u_i/\nu$ & $\rho(h_{\rm wm}/\mu)^2\tau$ & Eq. \eqref{eq:exp_linear} & F, F, F & F & F & {\color{blue}Y}\\
        \cite{zhou2021wall} & $\ln\left(\dfrac{h_{\rm wm}}{y^*}\right)$, $\dfrac{u_i}{U_b}\dfrac{\delta}{h_{\rm wm}}$, $\dfrac{h_{\rm wm}}{\rho U_b^2}\dfrac{\partial p}{\partial x_i}$ &$\dfrac{\tau_w}{\rho U_b^2}$ & Eq. \eqref{eq:relu} & -$\infty$, F, F & F & $-\infty$ & {\color{blue}N}\\
        \cite{zhou2023wall} & $\ln\left(\dfrac{h_{\rm wm}}{y^*}\right)$, $\dfrac{u_i}{U_b}\dfrac{\delta}{h_{\rm wm}}$, $\dfrac{h_{\rm wm}}{\rho U_b^2}\dfrac{\partial p}{\partial x_i}$ &$\dfrac{\tau_w}{\rho U_b^2}$ & Eq. \eqref{eq:tan_sigmoid} & $-\infty$, F, F & F & F & Y\\
    \hline
    \end{tabular}}
    \caption{Similar to the Table \ref{tab:WMLES}, while the test case is the laminar flow, i.e. $\rm Re_\tau\to 0$}
    \label{tab:WMLES-laminar}
\end{table}
\color{black}

\subsection{RANS models}
\label{sub:sgs}

There is a wealth of literature on data-enabled RANS models.
For brevity, we study the RANS models in Refs.\cite{singh2017machine, xie2021artificial,xiao2020flows}.
Details of these FNN-based models are shown in Table \ref{tab:RANS} including the references, the inputs and outputs, the activation functions, the behaviors of the inputs and outputs as the Reynolds number approaches infinity, as well as whether these FNNs preserve the log layer physics.

In Singh et al.\cite{singh2017machine}, an FNN is trained to augment the production term in the Spalart-Allmaras(SA) model. 
The FNN employed the sigmoid activation function.
The relevant inputs of the FNN are
\begin{equation}
\left\{ {\overline{\Omega}},\chi, \frac{S}{\Omega},\frac{\tau}{\tau_{w}},\frac{P}{D}, f_d\right\},
\label{eq:input_singh}
\end{equation}
where $\overline{\Omega}=y^2/(\hat{\nu}+\nu) \Omega$ is the mean vorticity magnitude normalized with the local scales, $\hat{\nu}$ is the SA working variable, $\chi=\hat{\nu}/\nu$, $S$ is the magnitude of the mean strain-rate tensor, $\tau$ is the magnitude of the Reynolds stress tensor, $\tau_{w}$ is the wall-shear stress, $P$ and $D$ are the production and destruction terms in the SA equation, and $f_d$ is a shielding function used in detached-eddy simulation\cite{spalart2006new}.
The inputs that are not relevant to the logarithmic layer are not detailed here for brevity.
As $\rm Re_\tau \to +\infty$, 
we have $\overline{\Omega}=O(1)$, $\chi =O(\rm Re_\tau)$, $S/\Omega=O(1)$, $\tau/\tau_w=O(1)$,  $P/D=O(1)$, and $f_d=O(1)$ (see Appendix \ref{sub:SA} for further details).
}
Hence, the expected network output is finite---SA requires no further augmentation to predict the flow in the logarithmic layer.
The actual output, according to Theorem \ref{th:multibound}, is also finite.
Hence, the model in Ref. \cite{singh2017machine}, if well trained, should preserve the log-layer physics.
The above argument applies equally to the machine-learning model in \cite{xiao2020flows}.
There, an FNN is trained to predict the error in the Reynolds stress tensor.
The baseline model is the $k$-$\omega$ model.
The model readily captures the log layer physics, and no further correction is needed.
That is, the expected output should be 0 at the infinite Reynolds number limit.
The is also the actual output at the infinite Reynolds number limit: at this limit, all inputs are finite and per Theorem \ref{th:multibound}, the output is also finite.

Xie et al. \cite{xie2021artificial} employed the velocity gradient and temperature gradient as inputs to their leaky-ReLu-activated FNN. 
The outputs of their model are the Reynolds stresses $\tau_{ij}$ and turbulent heat flux $Q_j$. 
Both the inputs and outputs were nondimensionalized using their root-mean-square values.
By employing this normalization, Xie et al. \cite{xie2021artificial} ensured that even as the Reynolds number tends towards infinity, the inputs and outputs converge to finite values. 
Hence, their models also preserve the log layer physics.

\begin{table}[!htb]
    \centering
    \begin{tabular}{rccccccc}\hline
         Ref. & Input & Output & Activation & Input (Ex)  & Output (Ex) & Model (Ex) & Ability \\
         \hline
        \cite{singh2017machine} & Eq. \ref{eq:input_singh}                &       $\beta$     & Eq. \eqref{eq:sigmoid} & F, $\rm Re_\tau$,  F, F, F, F & F & F & Y\\
          \cite{xiao2020flows} &  Table 2 in \cite{xiao2020flows}   &  $\xi$, $\eta$      &  Eq. \eqref{eq:relu}       & F & F, F & F & Y\\
          \cite{xie2021artificial}&     $\frac{\partial u_i}{\partial x_j},~ \frac{\partial T}{\partial x_j}$           & $\tau_{ij},~Q_j$         &         Eq. \eqref{eq:relu_leak} & F & F & F & Y\\
    \hline
    \end{tabular}
    \caption{Details of the FNN-based RANS models.}
    \label{tab:RANS}
\end{table}

\section{Conclusions}
\label{sec:conclusions}

Assessing the compliance of a black-box machine learning model with known physics scalings requires re-training and re-interfacing with a CFD code and therefore is not straightforward.
The sheer abundance of machine learning models in the literature adds to the complexity, posing significant challenges to their validation, verification, and subsequent application in engineering practices.
This paper aims to provide a solution to this practical challenge.
We develop mathematical frameworks that allow {\it a priori} screening of FNN-based machine-learning models for their compliance with known physics scalings. 
The Theorems in Section \ref{sub:theorem} are invoked to screen FNN-based wall models and RANS models for their compliance with the law of the wall, the log-layer physics, and the low Reynolds number physics.
The analysis shows that some FNN-based models preserve the law of the wall while others fall short.
It is important to note that the presented theorems provide necessary conditions. 
Consequently, FNNs identified as potentially preserving the log law must be meticulously trained to uphold the log law at high Reynolds numbers. 
On the other hand, FNNs identified as incapable of preserving the log law will inevitably fail to do so, regardless of the training method employed. 
This assertion carries significant weight.
Although {\it a posteriori} tests of the models are not pursued in this study, the conclusions here align well with the observations in Ref. \cite{vadrot2022survey}, lending further validation.
In that paper, the authors implemented the wall models in Refs. \cite{yang2019predictive,zhou2021wall,zhou2023wall,bae2022scientific} in an LES code and assessed their compliance with the law of the wall at Reynolds numbers from $Re_\tau=180$ and $10^{10}$. 
Lastly, it is worth noting that the theorems presented in this work not only aid in validating machine learning models but also offer guidance for network design. Referring back to the example in Section \ref{sub:theorem}, if one were to train a network to predict the centerline velocity in a channel based on the friction Reynolds numbers, the inputs and outputs should be designed to ensure that the expected and actual outputs exhibit the same asymptotic behavior.

\section*{Funding Sources}

Chen, Bin, and Shi are supported by NCSF grant number 91752202.
Yang is supported by the Office of Naval Research contract N000142012315 and the Air Force Office of Scientific Research award number FA9550-23-1-0272.
Abkar is supported by the Independent Research Fund Denmark (DFF) under the Grant No. 1051-00015B.

\section*{Acknowledgments}

Yang acknowledges George Huang for the fruitful discussion.
Chen acknowledges Jiaqi Li and Xinyi Huang for their constructive discussion of the machine-learning literature. 

\section{Appendix: Proofs of the theorems}
\label{sub:proof}
The proofs of the theorems in section \ref{sub:theorem} are straightforward, but 
they are provided here for completeness. 
The proof of Theorem \ref{th:singlebound} follows 
Ref. \cite{hornik1989multilayer}, where a FNN with a single hidden layer is expressed  as
\color{black}
\begin{equation}
    f(\mathbf{x})=\sum_{i=1}^q \beta_i \sigma(A_i(\mathbf{x})), ~\mathbf{x}\in \mathbb{R}^N,~\beta_i \in \mathbb{R},A_j \in A^N,~q \in \mathbb{N},
    \label{eq:single}
\end{equation}
where 
\begin{equation}
    A^N=\left\{f: \mathbb{R}^N\rightarrow \mathbb{R}: f(\mathbf{x})=\mathbf{\omega}\cdot \mathbf{x}+b, \mathbf{\omega}\in \mathbb{R}^N, b \in \mathbb{R} \right\}.
\end{equation}
We have
\begin{equation}
    |f(\mathbf{x})|=|\sum_{i=1}^q \beta_i \sigma(A_i(\mathbf{x}))|\le\sum_{i=1}^q |\beta_i| |\sigma(A_i(\mathbf{x}))|.
\end{equation}
$|\sigma|$ is bounded and therefore there exists a $c$ such that $|\sigma|<c$; and since $|\sigma|<c$, $|f(\mathbf{x})|< \sum\limits_{i=1}^q |\beta_i| c$, hence the boundedness of the FNN.

Theorem \ref{th:multibound} is proved as follows: 
\noindent
Let the activation function of the $s$th hidden layer be bounded.
If $s=r$, per Theorem 1, $f$ is bounded.
If $s<r$, per Theorem 1, the inputs to the $(s+1)$th hidden layer are finite.
Because the map from any hidden layer to the next is continuous, the output of any subsequent hidden layers must also be bounded.
Hence, $f$ is bounded.

Theorem \ref{th:singleunbound} is proved as follows:
\noindent
Define $x_m=\max\limits_{1\le i \le N} |x_i|$.
Per \cite{hornik1989multilayer}, we may write an FNN with a single hidden layer as Eq. \eqref{eq:single}.
Since $A_i(\mathbf{x})$'s are linear functions, $A_i(\mathbf{x})=O(x_m)$.
Consequently,
\begin{equation}
    \lim_{x_m\to \infty} \frac{|f(\mathbf{x})|}{x_m^\alpha}=\lim_{x_m\to \infty} \frac{|\sum_{i=1}^q \beta_i \sigma(A_i(\mathbf{x}))|}{x_m^\alpha}={\rm Const},
    \label{eq:proof3}
\end{equation}
Hence, $f(\mathbf{x})=O(x_m^\alpha)$.

Theorem \ref{th:multiunbound} is proved as follows:
\noindent
The output of $s$th hidden layer is given by
\begin{equation}
    x_{i,s}=\sigma\left(\sum_j \omega_{ij,s}x_{j,{s-1}}+b_{i,{s}}\right),
\end{equation}
where $\omega_{ij,s}$ and $b_{i,s}$ are the weights and biases of the $s$th hidden layer, respectively.
Similar to Eq. \eqref{eq:proof3}, we have
\begin{equation}
    \lim_{x_{m,s-1}\to \infty} \frac{x_{m,s}}{x_{m,s-1}^{\alpha_s}}=\lim_{x_{m,s-1}\to \infty} \frac{\max \left|\sigma\left(\sum\limits_j \omega_{ij,s}x_{j,{s-1}}+b_{i,{s}}\right)\right|}{x_{m,s-1}^{\alpha_s}}={\rm Const},
    \label{eq:proof4}
\end{equation}
where $x_{m,{s}}=\max_i |x_{i,s}|$. 
Hence, $x_{m,s}=O\left(x_{m,s-1}^{\alpha_s}\right)$, which, through the feedforward process, directly leads to theorem \ref{th:multiunbound}.

\section{Appendix: The details of the asymptotic behaviors of the SA models}
\label{sub:SA}
In this section, we show the details of the asymptotic behaviors of the SA models.
The one-equation Spalart–Allmaras turbulence model \cite{spalart1992one} solves for the modified eddy viscosity $\hat{\nu}$, which is defined as
\begin{equation}
    \hat{\nu}=\frac{\nu_t}{f_{v1}}, ~ f_{v1}=\frac{\chi^3}{\chi^3+c_{v1}^3}, ~ \chi=\frac{\hat{\nu}}{\nu}.
    \label{eq:nuhat}
\end{equation}
The model equation is as follows
\begin{equation}
    \frac{D \hat{\nu}}{D t}=P-D+\frac{1}{\sigma}\left[ \nabla\cdot \left(\left(\nu+\hat{\nu}\right)\nabla \hat{\nu}\right)+c_{b2}\left(\nabla \hat{\nu}\right)^2 \right].
    \label{eq:SA}
\end{equation}
The production and destruction terms are defined as follows
\begin{equation}
    P=c_{b1}\Tilde{\Omega}\hat{\nu},
    \label{eq:prod}
\end{equation}
\begin{equation}
    D=c_{w1}f_w\left[\frac{\hat{\nu}}{y}\right]^2.
    \label{eq:dest}
\end{equation}
Here
\begin{equation}
    \Tilde{\Omega}=\Omega+\frac{\hat{\nu}}{\kappa^2 y^2}f_{v2},
    \label{eq:omega-tilde}
\end{equation}
\begin{equation}
    f_{v2}=1-\frac{\chi}{1+\chi f_{v1}},
    \label{eq:fv2}
\end{equation}
\begin{equation}
    f_w=g\left[\frac{1+c_{w3}^6}{g^6+c_{w3}^6}\right]^{1/6},
    \label{eq:fw}
\end{equation}
\begin{equation}
    g=r+c_{w2}(r^6-r), ~ r=\frac{\hat{\nu}}{\Tilde{\Omega}\kappa^2y^2},
    \label{eq:g}
\end{equation}
$c_{b1}=0.1355$, $\sigma=2/3$, $c_{b2}=0.622$, $\kappa=0.41$, $c_{w1}=c_{b1}/\kappa^2+(1+c_{b2})/\sigma$, $c_{w2}=0.622$, $c_{w3}=2.0$, and $c_{v1}=7.1$ are constants.

Now we analysis the asymptotic behaviors of the input feature of the model in Singh et al. \cite{singh2017machine} as $\rm Re_\tau $ approaching infinity in the log layer of the canonical wall-bounded turbulent flows.
Firstly, the eddy viscosity $\nu_t=\kappa u_\tau y$.
Then 
\begin{equation}
    \chi=\frac{\nu_t}{f_{v1} \nu }=\frac{\kappa y}{\delta f_{v1}}\rm Re_\tau.
\end{equation}
Note that $f_{v1}=\chi^3/(\chi^3+c_{v1}^3)$.
Consequently, as $\rm Re_\tau \to +\infty$, $\chi=O(\rm Re_\tau)$, $f_{v1}\to 1$.
Then $\hat{\nu}\sim \kappa u_\tau y$, as $\rm Re_\tau \to +\infty$.
Here $A\sim B$ as $\rm Re_\tau \to +\infty$ means $\lim_{\rm Re_\tau \to +\infty}A/B=1$.

Since  as $\rm Re_\tau \to +\infty$, $\Omega\sim \partial U/\partial y=u_\tau/(\kappa y)$, 
\begin{equation}
    \lim_{\rm Re_\tau \to +\infty}\overline{\Omega}=\lim_{\rm Re_\tau \to +\infty} \frac{y^2}{\nu+\hat{\nu}}\Omega=\lim_{\rm Re_\tau \to +\infty}\frac{y^2}{\kappa y u_\tau}\frac{u_\tau}{\kappa y}=\kappa^{-2}.
\end{equation}
Similarly,
\begin{equation}
    \lim_{\rm Re_\tau \to +\infty}\tilde{\Omega}= \lim_{\rm Re_\tau \to +\infty}\left(\frac{u_\tau}{\kappa y}+\frac{\kappa y u_\tau}{\kappa^2 y^2}f_{v2}\right)= \lim_{\rm Re_\tau \to +\infty}\frac{u_\tau}{\kappa y}(1+f_{v2}).
\end{equation}
\begin{equation}
    \lim_{\rm Re_\tau \to +\infty} r=\lim_{\rm Re_\tau \to +\infty}\frac{\kappa y u_\tau }{\frac{u_\tau}{\kappa y}(1+f_{v2})\kappa^2 y^2}=\lim_{\rm Re_\tau \to +\infty}\frac{1}{1+f_{v2}}.
    \label{eq:r-infty}
\end{equation}
And
\begin{equation}
    \lim_{\rm Re_\tau \to +\infty}f_{v2}=\lim_{\rm Re_\tau \to +\infty}\left(1-\frac{1}{f_{v1}}\right)=0.
    \label{eq:fv2-infty}
\end{equation}
Consequently, $\lim_{\rm Re_\tau \to +\infty}f_w=\lim_{\rm Re_\tau \to +\infty}g=\lim_{\rm Re_\tau \to +\infty}r=1$,

Now we check the asymptotic behavior of $P/D$
\begin{equation}
    \lim_{\rm Re_\tau \to +\infty}\frac{P}{D}=\lim_{\rm Re_\tau \to +\infty}\frac{c_{b1} \Tilde{\Omega} \hat{\nu}}{c_{w1} f_w \left[ \frac{\hat{\nu}}{y}\right]^2}= c_1 \cdot \lim_{\rm Re_\tau \to +\infty}\frac{\frac{u_\tau}{\kappa y} \kappa u_\tau y}{(\kappa u_\tau)^2}=c_1 \kappa^{-2},
\end{equation}
where $c_1=c_{b1}/c_{w1}$.

The last input $f_d$ is a bounded function, $f_d\in [0, 2]$. 
And as $\rm Re_\tau \to +\infty$, $S/\Omega$ and $\tau/\tau_w$ should approach 1.

In summary, as $\rm Re_\tau \to +\infty$, 
we have $\overline{\Omega}=O(1)$, $\chi =O(\rm Re_\tau)$, $S/\Omega=O(1)$, $\tau/\tau_w=O(1)$,  $P/D=O(1)$, and $f_d=O(1)$.
\bibliographystyle{ieeetr}
\bibliography{a-v1}

\end{document}